\documentclass[%
reprint,
superscriptaddress,
frontmatterverbose,
 amsmath,amssymb,
prb,
]{revtex4-2}

\usepackage{amsfonts,amsmath,amssymb,amsthm}
\usepackage{graphicx}
\usepackage{bm}% bold maths

\usepackage{xcolor}
\definecolor{midnightblue}{cmyk}{1,1,0,0.1}
\definecolor{forestgreen}{cmyk}{0.76,0,0.26,0.5}

\usepackage{hyperref}
\hypersetup{
    unicode=false,          % non-Latin characters in Acrobat’s bookmarks
    pdftoolbar=true,        % show Acrobat toolbar?
    pdfmenubar=true,        % show Acrobat menu?
    pdffitwindow=false,     % window fit to page when opened
    pdfstartview={FitH},    % fits the width of the page to the window
    pdftitle={My title},    % title
    pdfauthor={Author},     % author
    pdfsubject={Subject},   % subject of the document
    pdfcreator={Creator},   % creator of the document
    pdfproducer={Producer}, % producer of the document
    pdfkeywords={keyword1} {key2} {key3}, % list of keywords
    pdfnewwindow=true,      % links in new window
    colorlinks=true,       % false: boxed links; true: colored links
    linkcolor=midnightblue,          % color of internal links (change box color with linkbordercolor)
    citecolor=magenta,        % color of links to bibliography
    filecolor=midnightblue,      % color of file links
    urlcolor=midnightblue,          % color of external links
}

\usepackage{ulem}
\usepackage{soul}
\usepackage{multirow}
\usepackage{makecell}
\usepackage{array}
\usepackage{threeparttable}

\begin{document}

\title{An implementation of the density functional perturbation theory\\ in the PAW framework}

\author{Xiaoqiang Liu}
\affiliation{International Center for Quantum Materials, School of Physics, Peking University, Beijing 100871, China}

\author{Yihao Lin}
\affiliation{International Center for Quantum Materials, School of Physics, Peking University, Beijing 100871, China}

\author{Ji Feng}\email{jfeng11@pku.edu.cn}
\affiliation{International Center for Quantum Materials, School of Physics, Peking University, Beijing 100871, China}
\affiliation{Hefei National Laboratory, Hefei 230088, China}
%\affiliation{Collaborative Innovation Center of Quantum Matter, Beijing 100871,
%China}
%\affiliation{CAS Center for Excellence in Topological Quantum Computation, University of Chinese Academy of Sciences, Beijing 100190, China}
\date{\today}% It is always \today, today,

\begin{abstract}
Quantifying materials' dynamical responses to external electromagnetic fields is central to understanding their physical properties. Here we present an implementation of the density functional perturbation theory for the computation of linear susceptibilities using the projector augmented-wave method. The Sternheimer equations are solved self-consistently through a nested iterative procedure to compute the first-order wavefunctions, from which the linear susceptibilities are obtained. As a demonstration, we compute the spin wave spectral functions of two magnetic metals. The computed magnon spectra for half-metallic CrO$_2$ and a Heusler intermetallic Cu$_2$MnAl show gapless Goldstone modes when spin rotation symmetry is preserved and display reasonable agreement with available experimental data. The Landau damping is computed to be small in CrO$_2$, but significant in Cu$_2$MnAl producing an asymmetric Lorentzian spectral lineshape. The access to linear susceptibilities as well as first-order wavefunctions offers a range of novel possibilities in quantitative understanding of materials' electronic properties from \textit{ab initio} methods.
\end{abstract}

\maketitle

\section{Introduction}

A microscopic understanding of electrical and magnetic characteristics of materials plays a key role in condensed matter physics, furnishing a unified insight into a wide range of phenomena. Indeed, the generalized density response functions (a.k.a. susceptibilities) of a many-electron system to external electromagnetic fields,~\cite{Kubo1957} in a broad sense encompasses the information of the usual longitudinal dielectric function and magnetic permeability, as well as the cross terms for electromagnetic coupling. The properties of collective excitations, e.g. plasmon and magnon, can also be procured from the susceptibilities. Therefore, computing the susceptibilities, which involve both charge and spin degrees of freedom of electrons, is essential to a full characterization of electronic properties. Though relatively straightforward for a non-interacting system, computing the susceptibilities for an interacting many-electron system is a nontrivial task due to interaction effects.

The Kohn-Sham density-functional theory (DFT)~\cite{Kohn1965} is by far the most widely employed ground state electronic structure method for materials and molecules. By mapping the ground state energy to a non-interacting system described by Kohn-Sham equations, the Kohn-Sham ansatz enables the variational formulation of the static responses of a many-electron system. The time-dependent (td) DFT~\cite{Runge1984} is subsequently developed, in which the electrodynamics is described by td Kohn-Sham equations. In these theories, the Hartree and exchange-correlation potentials are functionals of density, and treated as self-consistent fields. The self-consistent first-order perturbation in td DFT leads to the density functional perturbation theory (DFPT),~\cite{Gross1985} which is then the machinery for linear response calculations leading directly to the full response functions.

DFPT-based methods have been successfully applied to calculate the dielectric function~\cite{Hybertsen1987,Gajdo2006} and phonon dispersion,~\cite{Giannozzi1991,Baroni2001} with the results in quantitative agreement with experimental observations. In the computations of dielectric function, the first-order wavefunctions with respect to $\bm k$ are solved, while deformation potential from frozen atomic displacements is used as the external field to compute phonon dispersions. Dynamical responses to external electromagnetic fields from DFPT, which account for the screening of both charge and spin, have attracted considerable interest recently.~\cite{Savrasov1998,Buczek2009,Rousseau2012,Cao2018, Gorni2018, Dejean2020, Singh2020, Skovhus2021} In addition to quantifying electrical and magnetic properties of materials, the linear susceptibilities computed from DFPT also find applications in the $GW$ approximations.~\cite{Hybertsen1986,Shishkin2006,Shishkin2007}

Approaches to the DFPT can be broadly grouped into two categories: a Dyson-like equation is solved in the first,~\cite{Buczek2009,Rousseau2012,Skovhus2021} whereas the Sternheimer equations are solved in the second.~\cite{Savrasov1998,Cao2018,Gorni2018,Dejean2020,Singh2020} The Dyson-like equation approach starts with response functions computed for the Kohn-Sham ground state. Though formally transparent and amenable to various iterative techniques, the Dyson-like equation approach suffers from two shortcomings. It requires a large number of unoccupied states and huge planewave bases for adequate convergence.~\cite{Skovhus2021} More serious is the subtle basis set incompatibility between the Kohn-Sham states and the DFPT process, which gives rise to an artifactual spin excitation gap in systems with spin rotation symmetry. The latter problem can only be partially mended with delicate engineering of the interaction kernel.~\cite{Rousseau2012,Skovhus2022}
In the second category, the first-order wavefunctions are procured (often iteratively) by solving the Sternheimer equations, from which the density is updated  with charge mixing iterations and the response functions are computed upon convergence. In this case, one is forced to deal with wavefunctions and various pseudopotentials with all the technicalities,~\cite{Hamann1979,Vanderbilt1990,Blochl1994,Kresse1999} in addition to the nested iterative procedure. Since the full Kohn-Sham response functions are never required, this method is free of the burden of summation over a huge number of empty states. In addition, the first-order wavefunctions and densities computed are actually bonus, which can be useful for computing a variety of properties.

A particularly popular planewave-based approach to DFT is based on the projector augmented-wave (PAW) method.~\cite{Blochl1994,Kresse1999} Combining the formal simplicity of pseudopotentials and the versatility of the linearized augmented-planewave method, the PAW method offers both efficiency and accuracy to Kohn-Sham DFT calculations on extended solids, and a wide range of capabilities in various implementations.~\cite{Kresse1996,Giannozzi2017,Gonze2020} Despite its popularity, DFPT in the PAW framework has remained to be developed, which is accomplished in this work by solving the td Sternheimer equations to compute the linear susceptibilities of crystalline materials accounting for both charge and spin degrees of freedom. The paper is organized as follows. In Sec. \ref{sec:dfpt} the general theory of DFPT is introduced, from the viewpoint of the dressed spin in td external electromagnetic fields. The screening built on the notion of dressed spin leads to the Sternheimer equations in the frequency and momentum domain and explicit formula for the linear susceptibilities. In Sec. \ref{sec:dfptpaw} the PAW method is reviewed, based on which the formulation of DFPT in the PAW framework is described, along with a few implementation details. As a first calibration our implementation, the spin wave spectral functions are extracted from the computed linear susceptibilities. Two examples are presented: half-metallic CrO$_2$ (Sec. \ref{sec:cro2}) shows a clean spin wave spectrum and minimal Landau damping, and a full Heusler intermetallic Cu$_2$MnAl (Sec. \ref{sec:cu2mnal}) shows significant Landau damping in the spin excitations that can be quantified with a simple asymmetric Lorentzian lineshape. Lastly, a summary is provided with an eye on room for development from algorithm and physics points of view.

\section{Theory and implementation}
\subsection{Density functional perturbation theory}
\label{sec:dfpt}
The td DFT offers an efficient description of the dynamics of an interacting many-electron system in the presence of external fields.~\cite{Runge1984} As a self-consistent perturbation theory of td DFT, DFPT is introduced in this section, wherein the Sternheimer equations are specialized to crystalline systems under a monochromatic, periodic external electromagnetic field.

In td DFT, to account for both charge and spin degrees of freedom, the generalized density is given by
\begin{equation}
    \begin{aligned}
        \label{eq:dens}
\rho(\bm r,t) &= \sum_n \theta_n \operatorname{tr}\{  \psi_n^\dagger(\bm r,t)\sigma\psi_n(\bm r,t)\}\\
&=(\rho_0,\bm m)=(\rho_0,\rho_1,\rho_2,\rho_3)
    \end{aligned}
% =
% \begin{bmatrix}
% \varrho_{\uparrow\uparrow} & \varrho_{\uparrow\downarrow} \\
% \varrho_{\uparrow\downarrow} & \varrho_{\downarrow\downarrow}
% \end{bmatrix},
\end{equation}
where $\theta_n$ is the occupancy of the spinor single-particle state $\psi_n$, and the four-vector spin $\sigma = (\sigma_0,\sigma_1,\sigma_2, \sigma_3)$ ($\sigma_0$ is the  identity matrix and $\sigma_\alpha$ with $\alpha= 1,2,3$ are the Pauli matrices). The atomic units~\footnote[1]{$e=\hbar=m_e =1$} are adopted in this paper, so $\rho_0$ is the total charge density, and $\bm m$ is magnetization density. The dynamics of $\psi_n$  is prescribed by the td Kohn-Sham equations~\cite{Runge1984}
\begin{equation}
\mathrm{i} \partial_t |\psi_n(t)\rangle = [H + \delta H(t)]|\psi_n(t)\rangle.
\label{eq:tddft}
\end{equation}
The ground state Kohn-Sham Hamiltonian~\cite{Kohn1965} in Eq. (\ref{eq:tddft}) is
$
    H=-\frac{1}{2}\nabla^2+v[\rho^{(0)}](\bm r)
$
where the self-consistent potential is a functional of the ground state density $\rho^{(0)}$, composed of ionic, Hartree and exchange-correlation (xc) potentials, namely, $v^{\text i}$, $v^{\text H}$ and $v^{\text{xc}}$. Though the xc potential $v^{\text{xc}}$ as a functional of density is in principle nonlocal in space and time,~\cite{Vignale1996,Vignale1997, Kootstra2000,vanFaassen2002,Marques2006} the commonly adopted local and adiabatic approximation (ALDA) is assumed in this work.~\cite{Ando1977, Zangwill1980, Zangwill1980b}

The first-order Hamiltonian $\delta H$ comprises two contributions. The first arises from the coupling of the four-vector spin $\sigma$ with external fields 
\begin{equation}
    v^{\text{ext}}(\bm r,t) = -B_\alpha(\bm r, t)\sigma_\alpha,
\end{equation}
where the four-vector electromagnetic field $B(\bm r,t)=(-\phi,\frac{1}{2}\bm B)$.~\footnote[2]{Here, $\bm B$ is the Zeeman field. We ignore the nonlocal coupling between the magnetic field and orbitals, which requires the current density functional theory and is beyond the scope of the present work.} The indices $\alpha,\beta=0,1,2,3$ are implicitly summed over when repeated, but we will keep other summations explicit. The self-consistent inclusion of the density dependence in the Hartree and xc potentials means that $\delta H$ also includes a second contribution from induced density $\delta \rho(\bm r, t)$ that screens $v^{\text{ext}}$. In the adiabatic linear response theory, this can be formulated in terms of a \emph{dressed spin} $\tau_\alpha$
\begin{multline}
    \tau_\alpha(\bm r, \bm r',t-t') = -\frac{\delta H(\bm r, t)}{\delta B_\alpha(\bm r',t')}
   = \sigma_\alpha \delta(\bm r-\bm r')\delta(t-t')
    \\ 
   - \sigma_\gamma\int  f_{\gamma\beta}(\bm r, \bm r'') \chi_{\beta\alpha}(\bm r'', \bm r', t-t') \text d\bm r''.
\end{multline}
Here $\chi_{\alpha\beta}(\bm r,\bm r',t)$ is the linear susceptibility that we pursue in this work, defined via
\begin{equation}
    \delta\rho_\alpha(\bm r,t) = \int\! \chi_{\alpha\beta}(\bm r,\bm r',t-t')B_\beta(\bm r',t')\text d\bm r'\text dt'.
\end{equation}
The interaction kernel has two components, $f_{\alpha\beta}=f^{\text H}_{\alpha\beta}+f^{\text {xc}}_{\alpha\beta}$, namely, the Hartree and xc kernels in the ALDA 
\begin{equation}
	f_{\alpha\beta}(\bm r,\bm r') = \frac{\delta_{\alpha0}\delta_{\beta0}}{|\bm r\!-\!\bm r'|}
    +\frac{1}{2}\delta(\bm r-\bm r')
    \operatorname{tr}\left[\sigma_\alpha\frac{\partial v^{\text {xc}}}{\partial \rho_\beta}\right].
\end{equation}
In terms of the dressed spin, the first-order Hamiltonian in unscreened external fields is written 
\begin{equation}
    \delta H (\bm r , t )= -\int\! \tau_\alpha(\bm r, \bm r', t-t') B_\alpha(\bm r' , t') \text d\bm r'\text dt'.
    \label{eq:deltaH}
\end{equation}

Now with the first-order Hamiltonian, the first-order wavefunctions can be obtained by solving the Sternheimer equations~\cite{Sternheimer1954, deGironcoli1995, Savrasov1998, Baroni2001, Giustino2010, Cao2018}
\begin{equation}
	(\mathrm i\partial_t-H)|\psi^{(1)}_{n}(t)\rangle = \delta H(t)|\psi_{n}^{(0)}(t)\rangle,
\end{equation}
in which $\psi_{n}^{(\ell)}(t)$ is the $\ell$th-order wavefunction, with $|\psi_{n}^{(0)}(t)\rangle = e^{-\text i \varepsilon_{n}t }|\psi_{n}\rangle$. With the first-order wavefunctions, we will be able to compute the induced density via the variation of Eq. (\ref{eq:dens}), from which the first-order Hamiltonian will be updated.

\begin{figure}
    \includegraphics[width=70 mm]{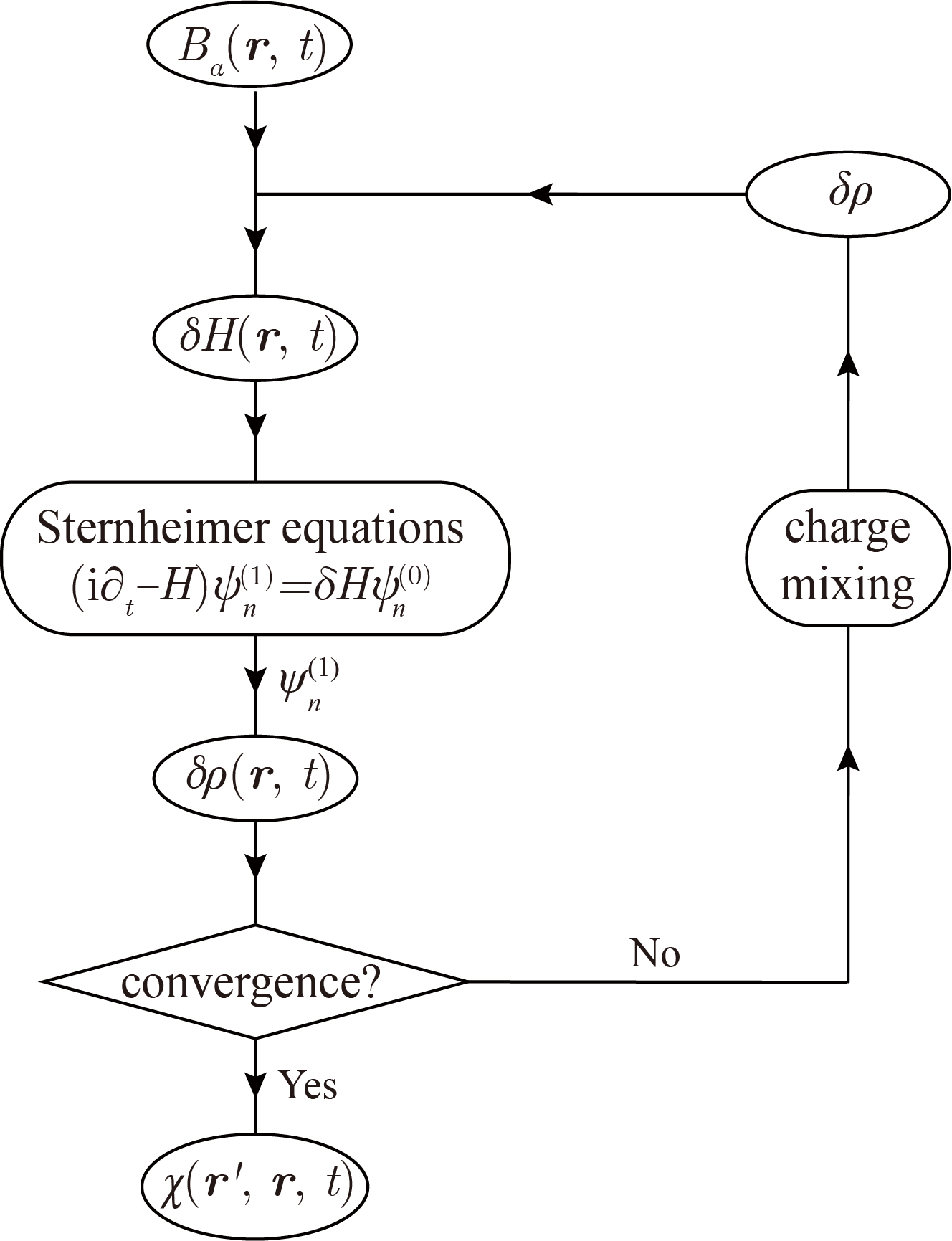}
    \caption{Flow chart for a nested-loop DFPT calculation in the Sternheimer equation approach. The outer loop depicted here is the charge mixing. The inner loop is incurred in solving the Sternheimer equations in each step of the outer iteration.}
    \label{fig:flow}
\end{figure}

With the above introduction, a DFPT calculation can then be performed in a nested iterative process depicted in Fig. \ref{fig:flow}. In the initializing step, $\delta H$ is constructed from the external electromagnetic fields, whence the Sternheimer equations  are solved in the inner iteration. This  produces a set of first-order wavefunctions, from which the induced density is calculated. A charge mixing strategy~\cite{Broyden1965,Johnson1988} is employed to revise the induced density, with the linear susceptibility and dressed spin computed subsequently. Then a new $\delta H$ is constructed from the updated spin to enter a second round of the outer iteration. The above process is repeated till convergence.

Now we will explain how the Sternheimer equations are solved in a crystalline solid. For electrons in a crystal, the initial Kohn-Sham states are Bloch functions such that 
$
    |\psi_n\rangle \mapsto |\psi_{n\bm k}\rangle =e^{\text i\bm k\cdot r} |u_{n\bm k}\rangle,
$
where $n$ is the band index and $\bm k$ is the crystal momentum, and $u_{n\bm k}$ is the cell-periodic part of the Bloch function.
The system is subject to spatially periodic and monochromatic external electromagnetic fields
\begin{equation}
	B_\alpha(\bm r, t) = \mathcal B_\alpha e^{\mathrm i(\bm p\cdot \bm r-\omega t)}+\mathrm{c.c.},
	\label{eq:Bext}
\end{equation}
where $\bm p=\bm q+\bm g$ with $\bm g$ being a reciprocal lattice vector for $\bm q$ to in the first Brillouin zone.
In this case, the following expansion is useful
\begin{equation}
\zeta(\bm r, t)=\sum_{l} e^{\mathrm i(\bm l\cdot \bm r-\nu t)}\zeta(\bm r, l)
 \label{eq:qexpand}
\end{equation}
for $\zeta=\delta H,\delta\rho_\alpha$, in which $ \zeta(\bm r,l)$ is a cell-periodic function with 
$$
l\equiv(\nu, \bm l) = \pm (\omega, \bm q).
$$
Then the first-order Hamiltonian in the $l$ channel is
\begin{equation}
    \delta H (\bm r , l)= - e^{-\mathrm i\bm l\cdot \bm r}\tau_\alpha(\bm r,\operatorname{sign}(\nu\omega)\bm p, \nu) \mathcal B_\alpha(l),
\end{equation}
where $\mathcal B_\alpha(\omega,\bm q)=\mathcal B_\alpha$, $\mathcal B_\alpha(-\omega,-\bm q)=(\mathcal B_\alpha)^*$, and
\begin{equation}
    \tau_\alpha(\bm r, \bm p', \nu) = \int \tau_\alpha(\bm r, \bm r',t) e^{\mathrm i (\bm p'\cdot \bm r'+\nu t)}\text d\bm r' \text dt.
\end{equation}

The first-order wavefunction is expanded as
\begin{eqnarray}
	|\psi_{n\bm k}^{(1)}(t)\rangle
	=\int\frac{\text d\nu}{2\pi} e^{-\mathrm i(\nu+\varepsilon_{n\bm k})t}|\psi^{(1)}_{n\bm k}(\nu)\rangle. 
   \label{eq:wexpand}
\end{eqnarray}
For the fields in Eq. (\ref{eq:Bext}), $|\psi^{(1)}_{n\bm k}(\nu)\rangle$ is nonzero only when $\nu=\pm \omega $ and can be written as
\begin{eqnarray}
    |\psi_{n\bm k}^{(1)}(\nu)\rangle
    =e^{\mathrm i(\bm k+\bm l)\cdot \bm r}|u^{(1)}_{n\bm k}(l)\rangle.
\end{eqnarray}
Then the Sternheimer equations become
\begin{equation}
	\left(\nu+\mathrm{i}\eta+\varepsilon_{n\bm k}-H_{\bm{k}+\bm l} \right) |u_{n\bm k}^{(1)}(l)\rangle =\delta H(\bm r,l) |u_{n\bm{k}}\rangle,
    \label{eq:SE}
\end{equation}
in which $H_{\bm k} = e^{-\mathrm i\bm k\cdot \bm r} H e^{\mathrm i\bm k\cdot \bm r}$ is the Bloch Hamiltonian. A positive infinitesimal $\eta$ is introduced on the left-hand side as a convergence factor to embody the causality structure of the linear susceptibility. In actual calculations, $\eta$ takes finite values to ensure convergence with finite $\bm k$-mesh especially for metals and bestows finite broadening on spectral peaks.

From the first-order wavefunctions, we compute the first-order induced density as
\begin{widetext}
    \begin{equation}
        \delta \rho_\alpha(\bm r,l)
        =\sum_{n\bm k} \theta_{n\bm k} \operatorname{tr} \left\{ u_{n\bm k}^\dagger (\bm r) \sigma_\alpha u^{(1)}_{n\bm k} (\bm r, l)
        +  {u^{(1)}_{n\bm k}}^\dagger (\bm r, -l) \sigma_\alpha u_{n\bm k} (\bm r) 
        \right\}.
    \end{equation}
The linear susceptibility can be extracted from the Fourier components of $\delta \rho_\alpha(\bm r, l)$
\begin{equation*}
    \delta\rho_\alpha(\bm {g}',l) = \chi_{\alpha\beta}(\bm g',\operatorname{sign}(\nu\omega)\bm g, l)\mathcal B_\beta(l),
\end{equation*}
whence
\begin{equation}
\begin{aligned}
    \chi_{\alpha\beta}(\bm g',\bm g'', l)=&\int e^{-\mathrm i(\bm g'+\bm l)\cdot \bm r'}\chi_{\alpha\beta}(\bm r',\bm r'', \nu)  e^{\mathrm i(\bm g''+\bm l)\cdot \bm r''} \text d\bm r'\text d\bm r''.
\end{aligned}
\end{equation}

It is worth mentioning that the Sternheimer equations in (\ref{eq:SE}) admit the following formal solutions in the $l$ channel
\begin{equation}
    |u^{(1)}_{n\bm k}(l)\rangle = - \mathcal B_\alpha(l)\sum_{n'}  \frac{|u_{n'\bm k+\bm l}\rangle \tau_\alpha(\bm k, l)_{n'n}}{\nu +\varepsilon_{n\bm k}\!-\!\varepsilon_{n'\bm k+\bm l}+\mathrm{i}\eta},
\end{equation}
in which the dressed spin matrix element is
% \begin{equation}
%     \tau_\alpha(\bm k, l)_{n'n} =
%     \langle u_{n'\bm k+\bm l} |e^{-\mathrm i\bm l\cdot \bm r}\tau_\alpha(\bm r,\operatorname{sign}(\nu\omega)\bm p,\nu) |u_{n\bm k} \rangle.
% \end{equation}
\begin{equation}
    \tau_\alpha(\bm k, l)_{n'n} =\int
    u^\dagger_{n'\bm k+\bm l}(\bm r)e^{-\mathrm i\bm l\cdot \bm r}\tau_\alpha(\bm r,\operatorname{sign}(\nu\omega)\bm p,\nu) u_{n\bm k} (\bm r) \text d\bm r.
\end{equation}
Because it requires the summation over a large number of empty states, this formal solution is not used in practice.~\footnote[3]{As will be explained later, in our implementation the exact solution is used to quickly correct the contributions of occupied and a small number of low-lying particle states in the iterative solution of the Sternheimer equations.}
The screened susceptibility then has the same expression as the (bare) Kohn-Sham susceptibility, except that the (unscreened) external fields are now coupled to the dressed spin; that is,
\begin{equation}
    \chi_{\alpha\beta}(\bm g',\operatorname{sign}(\nu\omega)\bm g, l) = -\sum_{nn'\bm k} (\theta_{n\bm k}-\theta_{n'\bm k+\bm l})\frac{ \langle u_{n\bm k} |e^{-\mathrm i \bm g'\cdot \bm r}\sigma_\alpha|u_{n'\bm k+\bm q}\rangle \tau_\beta(\bm k, l)_{n'n}}{\nu +\varepsilon_{n\bm k}-\varepsilon_{n'\bm k+\bm l}+\mathrm{i}\eta}.
\end{equation}
In arriving at the last expression, we have used the fact that  $\tau_\alpha(\bm k, l)_{n'n}=\tau_\alpha(\bm k+\bm l, -l)^*_{nn'}$.
\end{widetext}

\subsection{DFPT with PAW method}
\label{sec:dfptpaw}
The previous Subsection presents a sketch of the DFPT for periodic systems without recourse to computational details. In practical calculations, however, various technologies have been developed such that one can perform DFT (and therefore DFPT) calculations on valence electrons only for crystalline materials using planewave bases with the aid of pseudopotentials, to yield satisfactory accuracy with high efficiencies. It is well known that the norm-conserving pseudopotential~\cite{Hamann1979} requires large planewave bases particularly for localized orbitals in transition elements, while the application of ultrasoft pseudopotential~\cite{Vanderbilt1990} is partly limited by the rather laborious construction. In contrast, the PAW method, combining the pseudopotential and linearized augmented-plane-wave methods, is free from the above difficulties and has been used widely. Thus, performing DFPT within the PAW framework~\cite{Blochl1994,Kresse1999} for inhomogeneous and td electromagnetic fields is evidently useful though notably nontrivial. DFPT with the PAW method has been implemented within the Vienna ab initio simulation package (VASP)~\cite{Kresse1996} for atomic displacements in the static and long-wavelength limit to calculate zone-center phonon energies. The extension to inhomogeneous and td electromagnetic fields is accomplished in this work based on VASP 5.4.4, and a few implementation details warrant further clarification.  Here, we will briefly review the PAW method and describe how it is used in our DFPT calculations. Although the formalism for the ground state quantities is identical to those in the literature and notationally notorious, we feel compelled to provide some of these details, particularly in view of the time- and position-dependent external fields involved in our implementation.

The PAW method is based on a linear transformation between the all-electron (AE) Hilbert space orthogonal to core states and pseudo (PS) Hilbert space. The AE and PS wavefunctions are related by
\begin{equation}
    |\psi_{n\bm k} \rangle = T |\tilde{\psi}_{n\bm k} \rangle,
\end{equation}
with the linear operator defined as
\begin{equation}
    T = 1 +\sum_i (|\phi_{i} \rangle -|\tilde{\phi}_{i}  \rangle)\langle \tilde{p}_{i}  |.
\end{equation}
The index $i$ is a shorthand encapsulating the atomic site located at $\bm R_i$ as well as the quantum numbers  $(nlm)$ of the local orbitals and spin. $\phi$, $\tilde{\phi}$ and $\tilde{p}$ are AE partial waves, PS partial waves and projector functions, respectively, and should all be understood as spinors. In order to perform the DFPT calculations on a crystalline solid in the electromagnetic fields in Eq. (\ref{eq:Bext}), the key is to find the cell-periodic part of each PAW-pseudized quantity, especially the nonlocal ones. 

Upon application of the time-independent transformation $T$ to the first-order wavefunctions
% \begin{equation}
%     |\psi_{n\bm k}^{(1)}(t) \rangle = T |\tilde{\psi}_{n\bm k}^{(1)}(t) \rangle
% \end{equation}
 the pseudized Sternheimer equations read
\begin{equation}
    (\mathrm i \partial_t S-\tilde{H})|\tilde{\psi}^{(1)}_{n\bm k}(t)\rangle = \delta \tilde{H}(t)|\tilde{\psi}_{n\bm k}^{(0)}(t)\rangle,
\end{equation}
with $S =  T^{\dagger}T$, $\tilde H=  T^{\dagger}HT$ and $\delta \tilde{H}(t)=T^\dagger \delta {H}(t)T$. 

According to the implementation of PAW method in VASP,~\cite{Kresse1999} we have
\begin{equation}
\begin{aligned}
S &= 1+\sum_{ij}\left|\tilde{p}_{i}\right\rangle q_{ij}\left\langle \tilde{p}_{j}\right|,\\
\tilde{H}&= -\frac{1}{2} \Delta + \tilde{v}^{\mathrm{eff}}+\sum_{ij}\left|\tilde{p}_{i}\right\rangle D_{ij}\left\langle \tilde{p}_{j}\right|,
\end{aligned}
\end{equation}
in which the nonlocal potential  $D_{ij}= \hat D_{ij}+\tilde D_{ij}$ with $\tilde D_{ij}=D^1_{ij}-\tilde D^1_{ij}$. The quantities $q_{ij}$, $D^1_{ij}$ and $\tilde D^1_{ij}$ are defined in  reference.~\cite{Kresse1999} Apparently, the local potential $\tilde{v}^{\mathrm{eff}}(\bm r)$ is a functional of pseudo density $\tilde n(\bm r)$ and compensation density $\hat n(\bm r) $, while $\tilde D_{ij}$ is a function of density matrix $\varrho$, i.e.
\begin{equation}
\label{eq:densitydepen}
\begin{aligned}
\tilde{v}^{\mathrm{eff}}= & \tilde{v}^{\mathrm{eff}}[\tilde n + \hat n ], \\
\hat D_{ij}= & \int\frac{1}{2}\operatorname{tr}\{\sigma_\alpha\tilde{v}^{\mathrm{eff}}(\bm{r})\} Q_{ij}^{\alpha}(\bm{r}) \text d\bm r,\\
\tilde D_{ij} =& \tilde D_{ij}(\varrho).
\end{aligned}
\end{equation}
We have hidden the functional dependence on the pseudized core densities in $\tilde{v}^{\mathrm{eff}}(\bm r)$, which are kept frozen during the DFPT calculations. $\tilde n(\bm r)$, $\hat n(\bm r)$ and $\varrho_{ij}$ are given by, respectively,
\begin{equation}
\begin{aligned}
\tilde n(\bm r)= & \sum_{n\bm{k}}\theta_{n\bm{k}}\operatorname{tr}\{\tilde{\psi}_{n\bm{k}}^\dagger(\bm r)\sigma\tilde{\psi}_{n\bm{k}}(\bm r)\}, \\
\hat n (\bm r)= & \sum_{i,j}\varrho_{ij}Q_{ij}(\bm{r}), \\
\varrho_{ij} =& \sum_{n\bm{k}}\theta_{n\bm{k}}\langle \tilde{\psi}_{n\bm{k}}|\tilde{p}_{i}\rangle \langle \tilde{p}_{j}|\tilde{\psi}_{n\bm{k}}\rangle.
\end{aligned}
\end{equation}
Here $Q_{ij}(\bm{r})=\sum_L \operatorname{tr}\{\sigma\hat Q_{ij}^{L}(\bm{r})\}$ with $\hat Q_{ij}^{L}(\bm{r})$ defined to construct the compensation density $\hat n(\bm r)$ in the reference.~\cite{Kresse1999} It should be noticed that for $\varrho_{ij}$, only elements with $\bm R_i = \bm R_j$ are useful in our calculations, while $D_{ij}$ are nonzero only when $\bm R_i = \bm R_j$.

Now we derive the expression for the first-order Hamiltonian $\delta \tilde{H}(t)$. For the fields in Eq. (\ref{eq:Bext}), the first-order local densities and potentials follow the same expansion as in Eq. (\ref{eq:qexpand}), while the first-order density matrix $\delta \varrho_{ij}$ and nonlocal potential $\delta D_{ij}$ can be expanded as
\begin{equation}
\delta \zeta_{ij}(t)= \sum_{l} e^{\mathrm i(\bm l\cdot \bm R_i-\nu t)}\delta \zeta_{ij}(l).
\end{equation}
Here, the factor $e^{\mathrm i\bm l\cdot \bm R_i}$ in the $l$ channel is introduced such that $\delta \zeta_{ij}(l)$ is cell-periodic, i.e., $\delta \zeta_{ij}(l) = \delta \zeta_{i'j'}(l)$ if the positions of atomic site for $i,j$ and $i',j'$ differ by a lattice vector.

\begin{widetext}
The contribution of external electromagnetic fields in $\delta \tilde{H}(t)$ can be calculated directly
\begin{equation}
\label{eq:Hextpaw}
\begin{aligned}
\delta \tilde{H}^\text{ext}(\bm r, t) = &\sum_{l} e^{-\mathrm i\nu t} \left[ e^{\mathrm i\bm l\cdot \bm r} v^\text{ext}(\bm r,l)\right.  + \sum_{ij} \left. e^{\mathrm i\bm l\cdot \bm R_i} \left|\tilde{p}_{i}\right\rangle D^\text{ext}_{ij}(l)  \left\langle \tilde{p}_{j}\right|\right],
\end{aligned}
\end{equation}
with
\begin{equation}
\begin{aligned}
v^\text{ext}(\bm r,l)=& -\mathcal B_\alpha(l) \sigma_\alpha e^{\operatorname{sign}(\nu\omega)\mathrm i \bm g\cdot \bm r},\\
D^\text{ext}_{ij}(l)=& \langle \phi_{i}|e^{\mathrm i\bm l\cdot (\bm r-\bm R_i)} v^\text{ext}(\bm r,l)|\phi_{j}\rangle -\langle \tilde{\phi}_{i}|e^{\mathrm i\bm l\cdot (\bm r-\bm R_i)} v^\text{ext}(\bm r,l)|\tilde{\phi}_{j}\rangle.
\end{aligned}
\end{equation}
The contribution to $\delta \tilde{H}(t)$ from the induced densities has a similar expression as in Eq. (\ref{eq:Hextpaw}) and can be calculated via an explicit finite difference, as $\tilde{H}$ is a functional of $\tilde n$, $\hat n $ and $\rho_{ij}$.
The first-order densities are found to be
\begin{equation}
\label{eq:denspaw}
\begin{aligned}
\delta \tilde{n}(\bm r, l) =&\sum_{n\bm k} \theta_{n\bm k} \operatorname{tr} \{ {\tilde{u}_{n\bm k}}^\dagger (\bm r) \sigma \tilde{u}^{(1)}_{n\bm k} (\bm r, l)+ \tilde{u}^{(1),\dagger}_{n\bm k} (\bm r, -l) \sigma \tilde{u}_{n\bm k} (\bm r) \}, \\
\delta \hat{n}(\bm r, l)= & \sum_{i,j} e^{\mathrm i\bm l(\bm R_i-\bm r)} \delta \varrho_{ij}(l)Q_{ij}(\bm{r}),\\
\delta \varrho_{ij}(l)=&\sum_{n\bm{k}}\theta_{n\bm{k}} [ \langle \tilde{u}_{n\bm{k}}|\tilde{p}_{i\bm{k}}\rangle \langle \tilde{p}_{j\bm{k}+\bm{l}}|\tilde{u}_{n\bm{k}}^{(1)}(l)\rangle +\langle \tilde{u}_{n\bm{k}}^{(1)}(-l)|\tilde{p}_{i\bm{k}-\bm{l}}\rangle \langle \tilde{p}_{j\bm{k}}|\tilde{u}_{n\bm{k}}\rangle ],
\end{aligned}
\end{equation}
where we define $|\tilde{p}_{i\bm k} \rangle = e^{-\mathrm i\bm k\cdot (\bm r-\bm R_i)} |\tilde{p}_{i} \rangle$. To linear order in external fields, the first order effective local potential is given by
\begin{equation}
\label{eq:finite1}
\delta \tilde{v}^{\mathrm{eff}}(l)\approx\tilde{v}^{\mathrm{eff}}[\tilde n + \hat n + \delta \tilde{n}(l)+\delta \hat{n}(l)]-\tilde{v}^{\mathrm{eff}}[\tilde n + \hat n ].
\end{equation}
Similarly, the first-order nonlocal potentials can be approximated as
\begin{equation}
\begin{aligned}
\label{eq:finite2}
\delta \hat D_{ij}(l)\approx & \int e^{\mathrm i\bm l\cdot(\bm r-\bm R_i)} \frac{1}{2}\operatorname{tr}\{\sigma_\alpha\delta\tilde{v}^{\mathrm{eff}}(\bm r,l)\} Q_{ij}^{\alpha}(\bm{r}) \text d\bm r ,\\
\delta \tilde D_{ij} (l)\approx& \tilde D_{ij}(\varrho+\delta \varrho(l)) -\tilde D_{ij}(\varrho).
\end{aligned}
\end{equation}
Though introduced as forward differences in Eq.(\ref{eq:finite1}) and (\ref{eq:finite2}), these quantities are evaluated using 4th-order centered finite differences,  with a step length of a thousandth the density variables.

With the above results, the final Sternheimer equations become
\begin{equation}
\label{eq:SEPAW}
\left(\nu S_{\bm{k}+\bm l}+\varepsilon_{n\bm k}S_{\bm{k}+\bm l}-\tilde{H}_{\bm{k}+\bm l} \right) |\tilde{u}_{n\bm k}^{(1)}(l)\rangle =\delta \tilde{H}_{\bm{k}}(l)|\tilde{u}_{n\bm{k}}\rangle,
\end{equation}
with
\begin{equation}
\begin{aligned}
S_{\bm{k}+\bm l}&=  1+\sum_{ij}\left|\tilde{p}_{i\bm{k}+\bm{l}}\right\rangle q_{ij}\left\langle \tilde{p}_{j\bm{k}+\bm{l}}\right|,\\
\tilde{H}_{\bm{k}+\bm l}&= -\frac{1}{2} \Delta_{\bm{k}+\bm l} + \tilde{v}^{\mathrm{eff}}+\sum_{ij}\left|\tilde{p}_{i\bm{k}+\bm{l}}\right\rangle D_{ij}\left\langle \tilde{p}_{j\bm{k}+\bm{l}}\right| ,\\
\delta \tilde{H}_{\bm{k}}(l)& = v^\text{ext}(l)+\delta\tilde{v}^{\mathrm{eff}}(l)  + \sum_{ij}\left|\tilde{p}_{i\bm{k}+\bm{l}}\right\rangle [ D^\text{ext}_{ij}(l)+\delta \hat D_{ij}(l)+\delta \tilde D_{ij} (l)]\left\langle \tilde{p}_{j\bm{k}}\right|.
\end{aligned}
\end{equation}
Here $\tilde{u}_{n\bm{k}}$ and $\tilde{u}_{n\bm k}^{(1)}(l)$ are the cell-periodic parts of corresponding pseudo wavefunctions, respectively.

The pseudized Sternheimer equations in $\pm (\omega, \bm q)$ channels are solved separately in each iteration using a variant of residual minimization method with a direct inversion in the iterative subspace (RMM-DIIS),~\cite{Pulay1980,Wood1985} which is already implemented in VASP 5.4.4. The L\"owdin perturbation theory is also performed to correct the first-order wavefunctions in the subspace of occupied states and low-lying excitations to speed up convergence
\begin{equation}
|\tilde{u}^{(1)}_{n\bm k}(l)\rangle \rightarrow |\tilde{u}^{(1)}_{n\bm k}(l)\rangle- \sum_{n'} |\tilde{u}_{n'\bm k+\bm l}\rangle \langle \tilde{u}_{n'\bm k+\bm l}|S_{\bm{k}+\bm l}|\tilde{u}^{(1)}_{n\bm k}(l)\rangle + \sum_{n'}  \frac{|\tilde{u}_{n'\bm k+\bm l}\rangle \langle \tilde{u}_{n'\bm k+\bm l} |\delta \tilde{H} |\tilde{u}_{n\bm k} \rangle}{\nu +\varepsilon_{n\bm k}-\varepsilon_{n'\bm k+\bm l}}.
\end{equation}
\end{widetext}
In the last equation above, the summations on $n'$ run over the occupied bands plus a few empty bands.

Apparently, solving Eq. (\ref{eq:SEPAW}) requires a $\bm k$-grid supplemented by two additional grids shifted by $\pm \bm q$ when $\bm q$ itself is not on the $\bm k$-grid. Doing so, however, not only increases the computational burden but also, more seriously, obliterates the exact cancellation of the contribution of the occupied 
manifold to the density change due to a $\bm k$-grid discretization error. The latter can be easily avoided by employing a pair of grids with a $\bm q$ shift, which also reduces the calculation partly. Then Eq. (\ref{eq:SEPAW}) is solved on the $\bm k$-grid in $+\bm q$ channel, and on the $\bm k+\bm q$ grid in $- \bm q$ channel. It is observed that in this dual grid setup, the above cancellation is well preserved.

The xc potentials in ALDA are functionals of real-valued densities. Thus, calculating $\delta \tilde{v}^{\mathrm{eff}}(l)$ like in Eq. (\ref{eq:finite1}) requires caution as $\delta \tilde{n}(l)$ and $\delta \hat{n}(l)$ are usually complex. In fact, the real and imaginary part of $\delta \tilde{v}^{\mathrm{eff}}(l)$ are calculated separately,
\begin{equation}
\mathfrak{F}\delta \tilde{v}^{\mathrm{eff}}(l)\approx \tilde{v}^{\mathrm{eff}}[\tilde n + \hat n + \mathfrak{F}\delta \tilde{n}(l)+\mathfrak{F}\delta \hat{n}(l)]-\tilde{v}^{\mathrm{eff}}[\tilde n + \hat n ]
\end{equation}
where $\mathfrak{F} = \operatorname{Re}, \operatorname{Im}$ takes the real or imaginary part, respectively. In the case of nonlocal potential, $\delta \tilde D_{ij} (l)$ and $\delta \varrho_{ij}(l)$ are first decomposed into two independent Hermitian matrices (i.e. Hermitian part and anti-Hermitian part multiplied by $-\mathrm{i}$), and then finite differenced separately in an analogous fashion.

Symmetry reduction is also performed in our implementation, where the summation over $\bm k$ points in Eq. (\ref{eq:denspaw}) is restricted to the symmetry-irreducible part of the Brillouin zone. The symmetry group here is the subgroup of the magnetic group of the studied crystal in which the external electromagnetic fields in Eq. (\ref{eq:Bext}) is invariant.

\section{Application to spin-wave spectrum calculation}

Our implementation enables computing the linear susceptibilities $\chi_{\alpha\beta}$ with the self-consistent inclusion of the interaction kernel. Directly inverting $\chi_{\alpha\beta}$ yields the dielectric tensor, which is composed of the usual charge sector $\epsilon_{00}$, spin sector $\epsilon_{\alpha\beta}$, and the spin-charge sector  $\epsilon_{0\beta}$, each embodying unique physics. Computing $\chi_{\alpha\beta}$ then can have diverse applications in evaluating materials properties pertaining to both charge and spin fluctuations, or in subsequent many-body calculations beyond the Kohn-Sham mean fields. One immediate application that has received considerable attention is the calculation of spin-wave excitation.~\cite{Savrasov1998, Buczek2009, Rousseau2012, Cao2018, Gorni2018,Dejean2020,Singh2020, Skovhus2021} According to the fluctuation-dissipation theorem, the spin-spin correlation function, directly accessible by various spin-sensitive inelastic scattering probes,~\cite{Mook1973,Braicovich2009,Ament2011} is related to the imaginary part of the linear susceptibilities,
\begin{equation}
S_{\scriptscriptstyle +-}(\bm p,\omega)=\frac{\operatorname{Im}\chi_{\scriptscriptstyle +-}(\bm g,\bm g,\omega,\bm q)}{1-e^{-\hbar\omega/k_{\mathrm B} T}}.
\end{equation}

\begin{widetext}

\begin{table*}[htpb]
\centering
\setlength{\tabcolsep}{3mm}{
\begin{threeparttable}[b]
\caption{Reported implementations of DFPT for spin wave spectra calculations by solving the Sternheimer equations.}
\label{tb:lit}
\begin{tabular}{cccc}
\hline
\hline
Authors (Year) & Potential & Basis set & Software\\
\hline
\makecell[l]{Savrasov (1998)~\cite{Savrasov1998}} & Full potential & LMTO\tnote{1} & LMTO Magnons \\
\makecell[l]{Cao et al. (2018)~\cite{Cao2018}} & NCPP\tnote{2}, USPP\tnote{3}~ & Planewave  & QE\tnote{4} \\
\makecell[l]{Gorni et al. (2018)~\cite{Gorni2018}} & NCPP & Planewave & QE \\
\makecell[l]{Tancogne-Dejean et al. (2020)~\cite{Dejean2020}} & NCPP & Real space & Octopus \\
\makecell[l]{Singh et al. (2020)~\cite{Singh2020}} & Full potential & LAPW\tnote{5} & Elk \\
\hline
\hline
\end{tabular}
$^1$Linear muffin-tin orbital.
$^2$Norm-conserving pseudopotential.
$^3$Ultrasoft pseudopotential.
$^4$Quantum ESPRESSO.
$^5$Linearised augmented-plane wave.
\end{threeparttable}
}
\end{table*}
\end{widetext}

Although for magnetic systems dominated by local moments the magnon can be described effectively by localized spin models, this method is subject to debate when delocalization sets in, and ultimately of questionable validity for itinerant magnetism. In these latter cases, which include a wide range of magnetic materials, the DFPT route becomes invaluable for computing the spin wave spectra \textit{ab initio}. To our knowledge, there have been just a handful of works devoted to implementing DFPT scheme for this purpose by solving the Sternheimer equations, as summarized in Table \ref*{tb:lit}. In these efforts, implementations are limited to the full-potential,~\cite{Savrasov1998, Singh2020} or norm-conserving and ultrasoft pseudopotentials.~\cite{Cao2018, Gorni2018, Dejean2020}

In this section, we present an initial application of our implementation of DFPT in the PAW framework to the calculations of spin-wave spectra for a couple of magnetic materials. For ferromagnets with the spin polarized along $z$ direction, transverse magnetic field can be applied by choosing $\mathcal B = (0, 1, -\mathrm{i}, 0)$ in Eq. (\ref{eq:Bext}), from which $\chi_{\scriptscriptstyle +-}$ is calculated directly. In these cases, the only remaining symmetry operation keeping the crystal and the transverse magnetic field unchanged is identity transformation. Thus, there is no room for symmetry reduction. For notationaly convenience, we define $\chi_{\scriptscriptstyle +-}(\bm p,\omega)\equiv\chi_{\scriptscriptstyle +-}(\bm g,\bm g,\omega,\bm q)$ as $\bm p=\bm q+\bm g$.

\subsection{Half-metallic chromium dioxide}
\label{sec:cro2}

\begin{figure}
    \centering
    \includegraphics[width=70 mm]{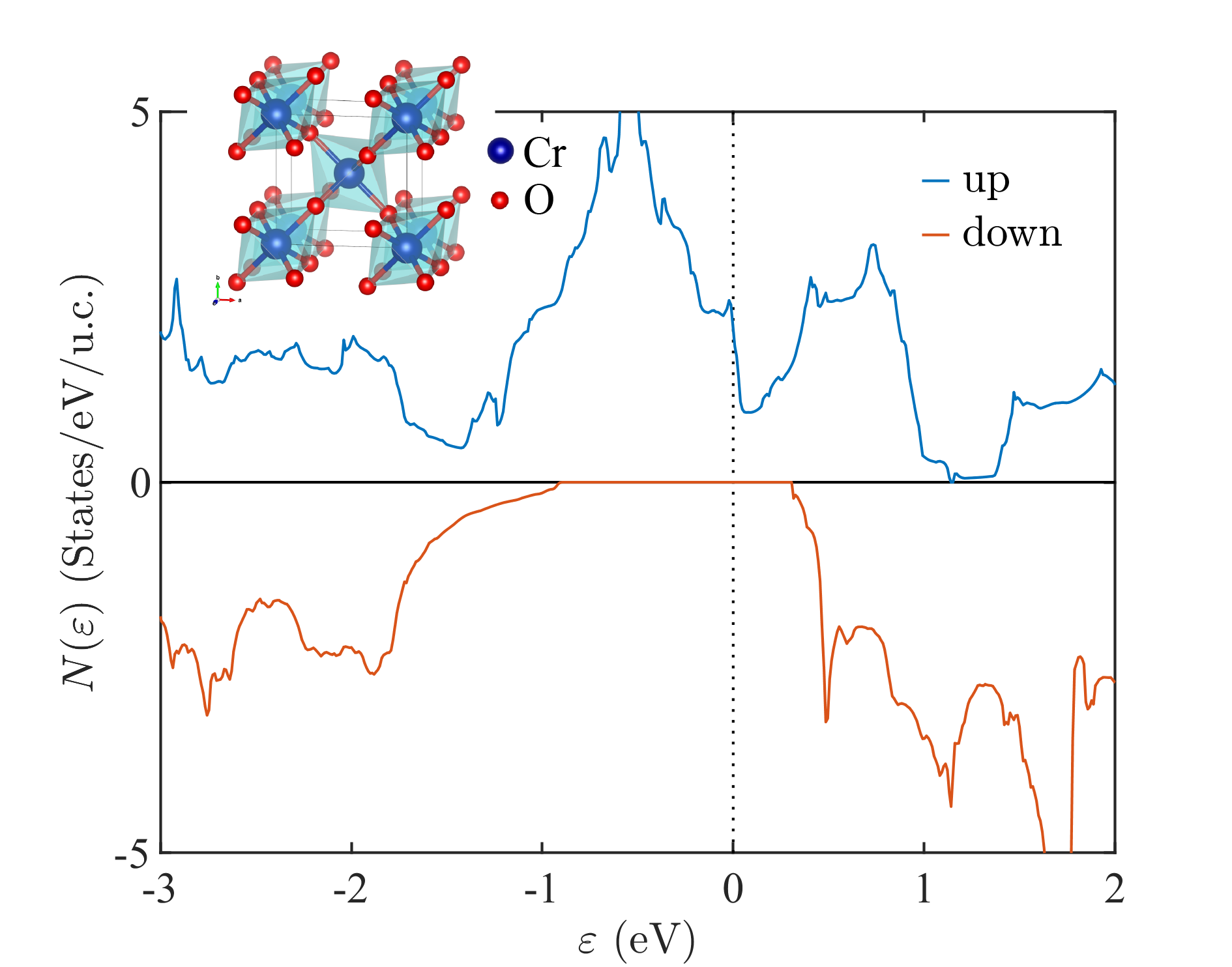}
    \caption{ Spin-resolved densities of states of CrO$_2$. The spin-flip gap is found to be about 310 meV. Inset: the crystal structure of CrO$_2$, highlighting the CrO$_6$ octahedra.}
    \label{fig:CrO2_struc}
\end{figure}

As shown in the inset in Fig. \ref{fig:CrO2_struc}, chromium dioxide, CrO$_2$, is a ferromagnetic half-metallic oxide with a rutile crystal structure, where each Cr atom is situated at the center of an octahedral cage formed by oxygen atoms.~\cite{Schwarz1986,Cloud1962}  Widely used as a magnetic recording material, CrO$_2$ also has various potential applications in spintronics and magnetoelectronics~\cite{Singh2016,Rabe2000} due to its half-metallic properties.

The experimental lattice parameters $a = b = 4.4218~$\AA~and $c = 2.9182~$\AA~\cite{Cloud1962} are used in our calculations. The planewave energy cutoff is set to be $500$ eV and a $13\times13\times20$ $\Gamma$-centered mesh of $\bm k$-points is used. The spin-resolved density of states of CrO$_2$ is computed from the collinear spin-polarized calculation and shown in Fig. \ref{fig:CrO2_struc}, where the half-metallic band structure is clearly seen. The magnetic moment of Cr is found to be 2 $\mu_{\mathrm B}$. We then turn to the noncollinear calculations, and compute the transverse spin susceptibility $\chi_{\scriptscriptstyle +-}(\bm p,\omega)$ along [100] and [001] directions for $\omega\leq 400$ meV on 10 meV intervals. The broadening parameter $\eta$ introduced in Eq. (\ref{eq:SE}) is set to be 50 meV in the calculations in this subsection.

\begin{figure}
    \centering
    \includegraphics[width=70 mm]{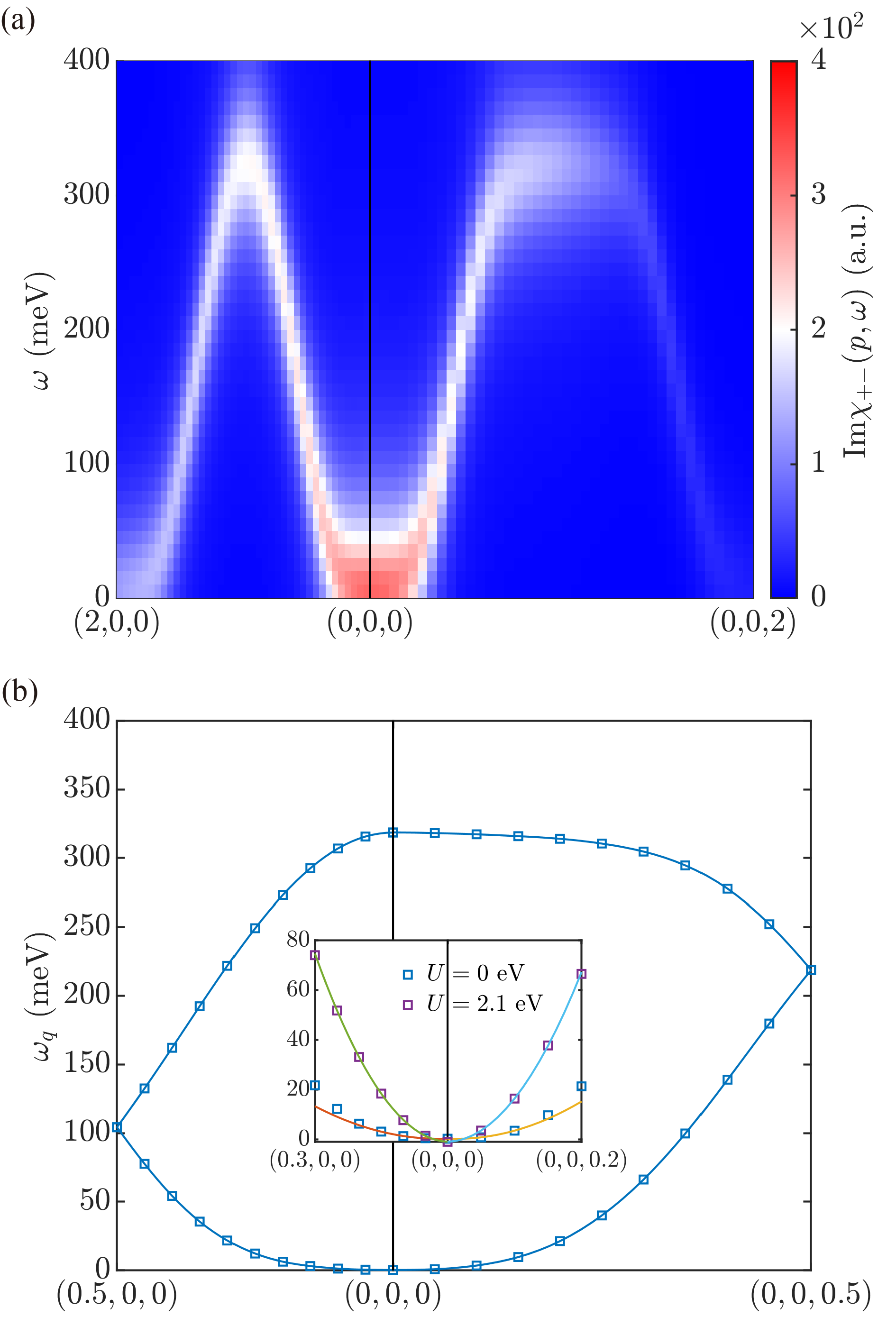}
    \caption{(a) Im$\chi_{\scriptscriptstyle +-}(\bm p,\omega)$ of CrO$_2$ along [100] and [001] directions. The black vertical line indicates the center of the first Brillouin zone. (b) The folded magnon energy dispersion $\omega_{\bm q}$ of CrO$_2$ in the first Brillouin zone, extracted from $\operatorname{Im}\chi_{\scriptscriptstyle +-}$ shown in (a). The inset shows the quadratic fits to $\omega_{\bm q}$ at small $q$ along [100] and [001] directions, respectively, for calculations without Hubbard $U$ correction and with $U^{\mathrm{eff}} = 2.1$ eV. The squares are the calculated data.}
    \label{fig:CrO2_Chi}
\end{figure}

Fig. \ref{fig:CrO2_Chi}(a) shows the computed Im$\chi_{\scriptscriptstyle +-}(\bm p,\omega)$, without spin-orbit interaction, along two $\bm p$ paths. In general, $\chi_{\scriptscriptstyle +-}(\bm p,\omega)$ is not periodic in $\bm p$. Then the branch in the first Brillouin Zone is composed of acoustic magnon modes, while the branch in the second Brillouin zone  optical modes. The profile of the magnon peak at a given $\bm p$ is nearly perfect Lorentzian over the entire energy range. The extracted half width at half maximum $\eta_{\bm p}$ are almost a constant and equal to the artificial broadening parameter $\eta$, indicating that the Landau damping in CrO$_2$ is negligible. This is expected given that half-metallic CrO$_2$ has a large spin-flip gap around 310 meV, as shown in Fig. \ref{fig:CrO2_struc}.

The location of the maxima of magnon peaks, $\omega_{\bm p}$, are then recorded and folded to the first Brillouin zone ($\omega_{\bm q}$). As shown in Fig. \ref{fig:CrO2_Chi}(b), we find one acoustic magnon branch and one optical branch, consistent with the fact that the unit cell in CrO$_2$ contains two magnetic Cr atoms. There is no magnon gap at the Brillouin zone boundaries, which is result of the $n$ glide symmetry. The energy of long-wavelength acoustic magnons is quadratic in $q$ with a gapless Goldstone mode, $\omega_{\bm q} = D_{\parallel}(q_x^2+q_y^2)+D_zq_z^2$, as expected for a ferromagnet with spin rotation symmetry. The spin stiffness coefficients are found to be $D_{\parallel} = 82 $ meV$\cdot$\AA $^2$ along [100] and $D_z = 92 $ meV$\cdot$\AA $^2$ along [001] directions, respectively. From this, we can estimate the average spin stiffness coefficient to be 85 meV$\cdot$\AA $^2$, which is close to the experimental measured result ($\sim 112.5$ meV$\cdot$\AA $^2$~\cite{Watts2002}).

For comparison, additional electron correlation is included statically within the LSDA+U formalism~\cite{dudarev1998} in both the ground state calculation and subsequent DFPT calculations, with $U^{\mathrm{eff}} = 2.1$ eV.~\cite{Korotin1998} As shown in the inset of Fig. \ref{fig:CrO2_Chi}(b), the gapless Goldstone mode is obtained again, accompanied by an average spin stiffness coefficient $D=391$ meV$\cdot$\AA $^2$, almost five times as large as the one without Hubbard $U$ correction. The energy of magnon in CrO$_2$ seems to be highly overestimated in LSDA+U calculations.
% $D_{\parallel} = 407 $ meV$\cdot$\AA $^2$ along [100] and $D_z = 361 $ meV$\cdot$\AA $^2$ along [001] directions

\begin{figure}
    \centering
    \includegraphics[width=70 mm]{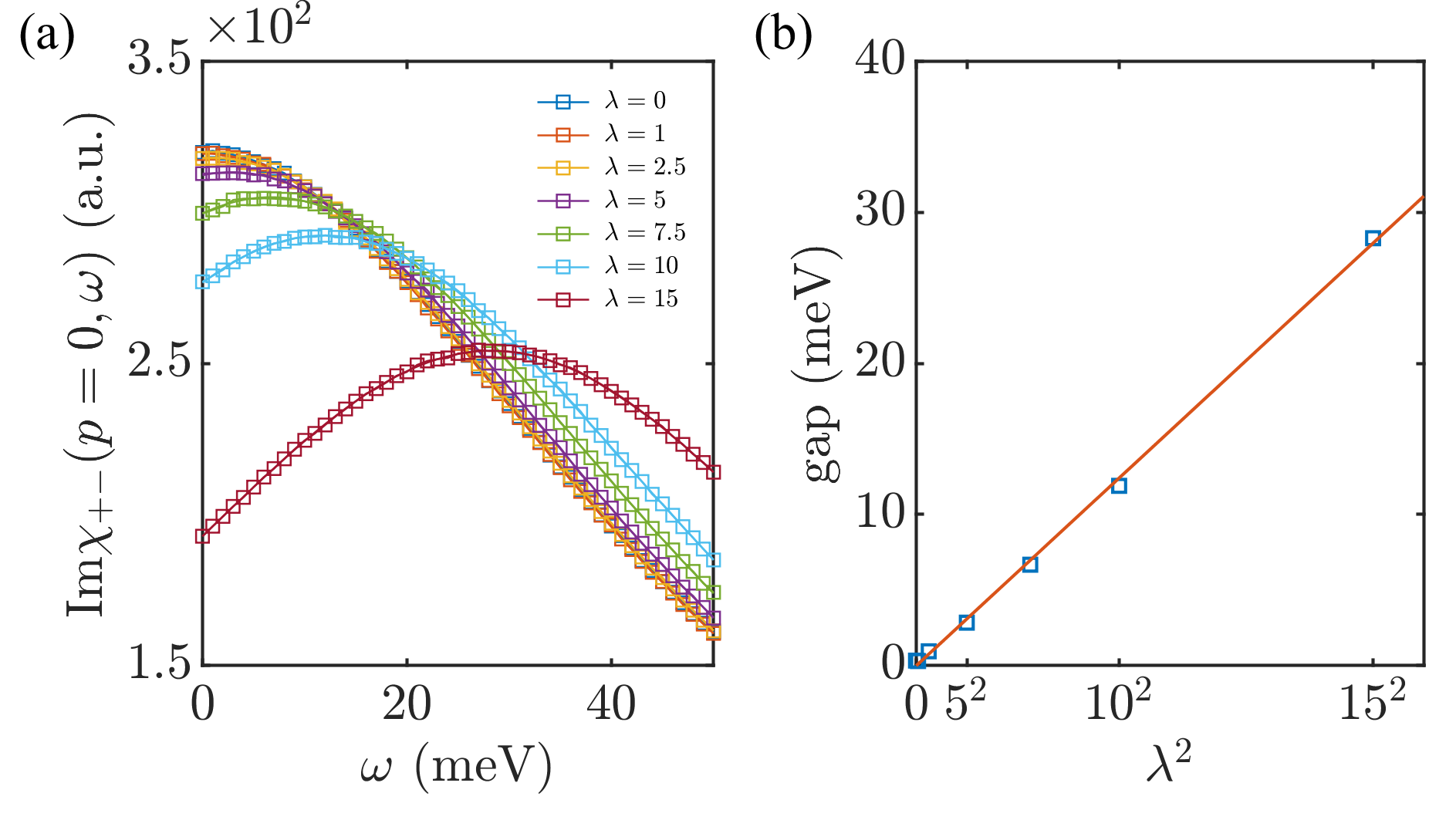}
    \caption{(a) Im$\chi_{\scriptscriptstyle +-}(p=0,\omega)$ as a function of $\omega$ for different $\lambda$ values. $\lambda = 1$ corresponds to the actual strength of spin-orbit coupling in CrO$_2$. The squares are the calculated data and the solid lines are the fits with Lorentzian line shape. (b) The Goldstone gap as a function of $\lambda^2$ in CrO$_2$. The solid line is the linear fit.}
    \label{fig:CrO2_Lambda}
\end{figure}

As a further test, we examine the Goldstone gap as a result of breaking the spin rotation symmetry by introducing spin-orbit interaction. The atomic spin-orbit interaction for Cr is fairly weak (on the order of tens of meV). Since the gap in the Goldstone magnon is second order in the spin-orbit coupling, it is small for CrO$_2$. In order to visualize  the effect of spin-orbit coupling, we introduce a parameter $\lambda$ to artificially tune its strength  (or speed of light), as in
$
H=H^0+\lambda H^{\mathrm{soc}},
$
where $\lambda = 1$ corresponds to the actual strength of spin-orbit coupling in CrO$_2$. The calculated Im$\chi_{\scriptscriptstyle +-}(p=0,\omega)$ as a function of $\omega$ for different $\lambda$ values are shown in Fig. \ref{fig:CrO2_Lambda}(a). Apparently there is blue shift of the Goldstone mode with increasing $\lambda$, indicating the emergence of a Goldstone gap. The Goldstone gap indeed shows a quadratic dependence on $\lambda$, as demonstrated by the gap-vs-$\lambda^2$ plot in Fig. \ref{fig:CrO2_Lambda}(b). The extrapolated Goldstone gap in CrO$_2$ is about 0.1 meV.

\subsection{Heusler intermetallic Cu$_2$MnAl}
\label{sec:cu2mnal}

\begin{figure}
    \centering
    \includegraphics[width=70 mm]{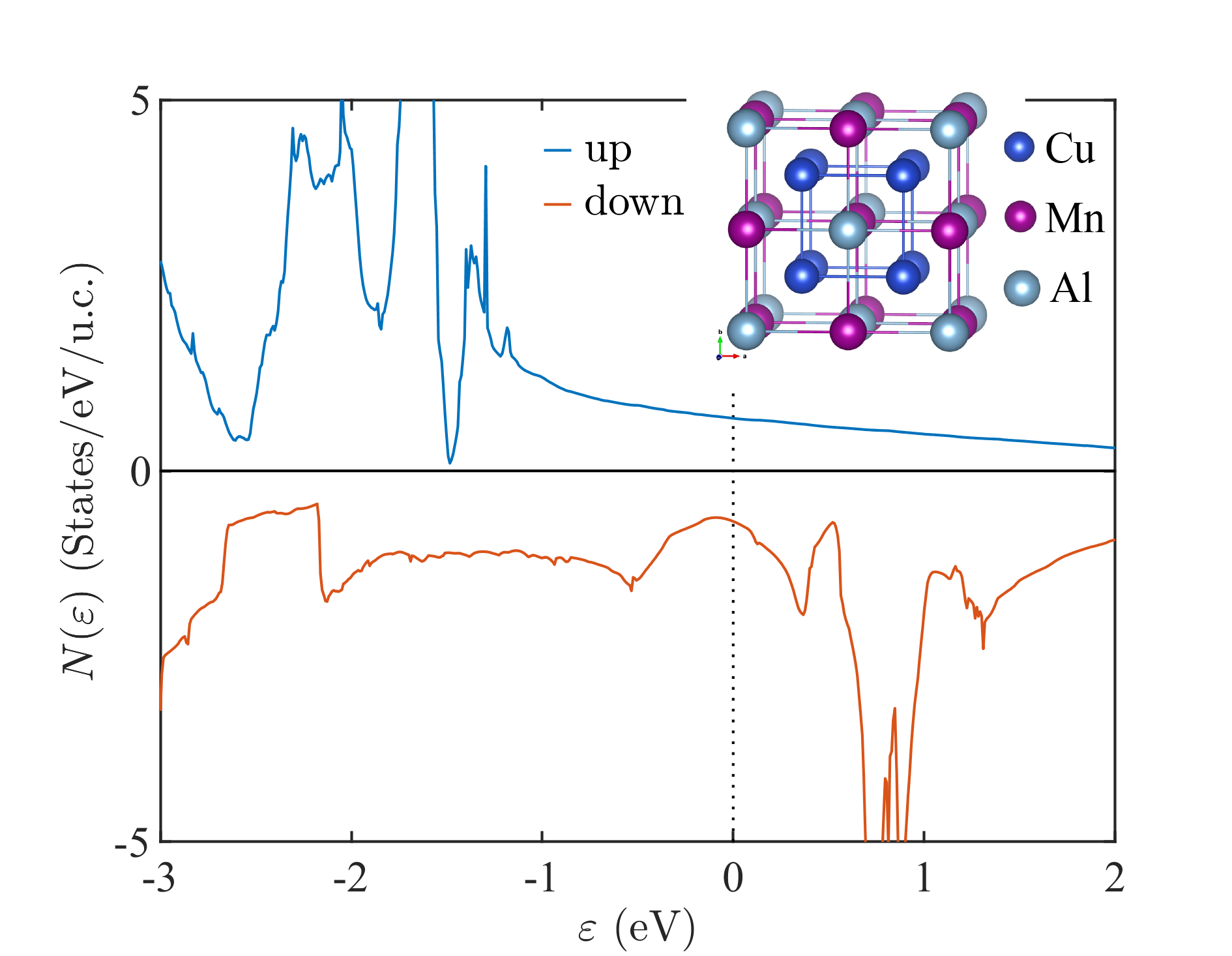}
    \caption{Spin-resolved densities of states of Cu$_2$MnAl. Inset: the crystal structure of full Heusler Cu$_2$MnAl, showing a conventional cubic unit cell for the $L2_1$ structure.}
    \label{fig:Cu2MnAl_struc}
\end{figure}

The ternary intermetallic Cu$_2$MnAl is a Mn-based full-Heusler alloy with the $L2_1$ structure type (see inset in Fig. \ref{fig:Cu2MnAl_struc}). The experimental lattice parameter for the conventional cubic cell (space group $Fm\bar 3m$) is $a = 5.968$~\AA.~\cite{Buschow1983} Cu$_2$MnAl is ferromagnetic below the relatively high Curie temperature (603 K).~\cite{Buschow1983} Apart from being regarded as a prototype for understanding the electronic correlations in Heusler intermetallics,~\cite{Weber2015} Cu$_2$MnAl  is also being used as a neutron polarizer and monochromator material.~\cite{Delapalme1971,Neubauer2012}

A planewave energy cutoff of $350$ eV and a $15\times15\times15$ $\Gamma$-centered  $\bm k$-grid are used in our calculations. Spin-orbit coupling is not included. The spin-resolved densities of states of Cu$_2$MnAl computed from the collinear spin-polarized calculation confirms the ferromagnetism of Cu$_2$MnAl, as shown in Fig. \ref{fig:Cu2MnAl_struc}. The magnetic moment is carried primarily by Mn atoms and computed to be 3.4 $\mu_{\mathrm B}$/Mn. The transverse spin susceptibility $\chi_{\scriptscriptstyle +-}(\bm p,\omega)$ along [100], [110] and [111] crystallographic directions is then computed in noncollinear calculations for $\omega\leq 300$ meV on 5 meV intervals, with a broadening parameter $\eta$ of 50 meV.

\begin{figure}[t]
    \centering
    \includegraphics[width=80 mm]{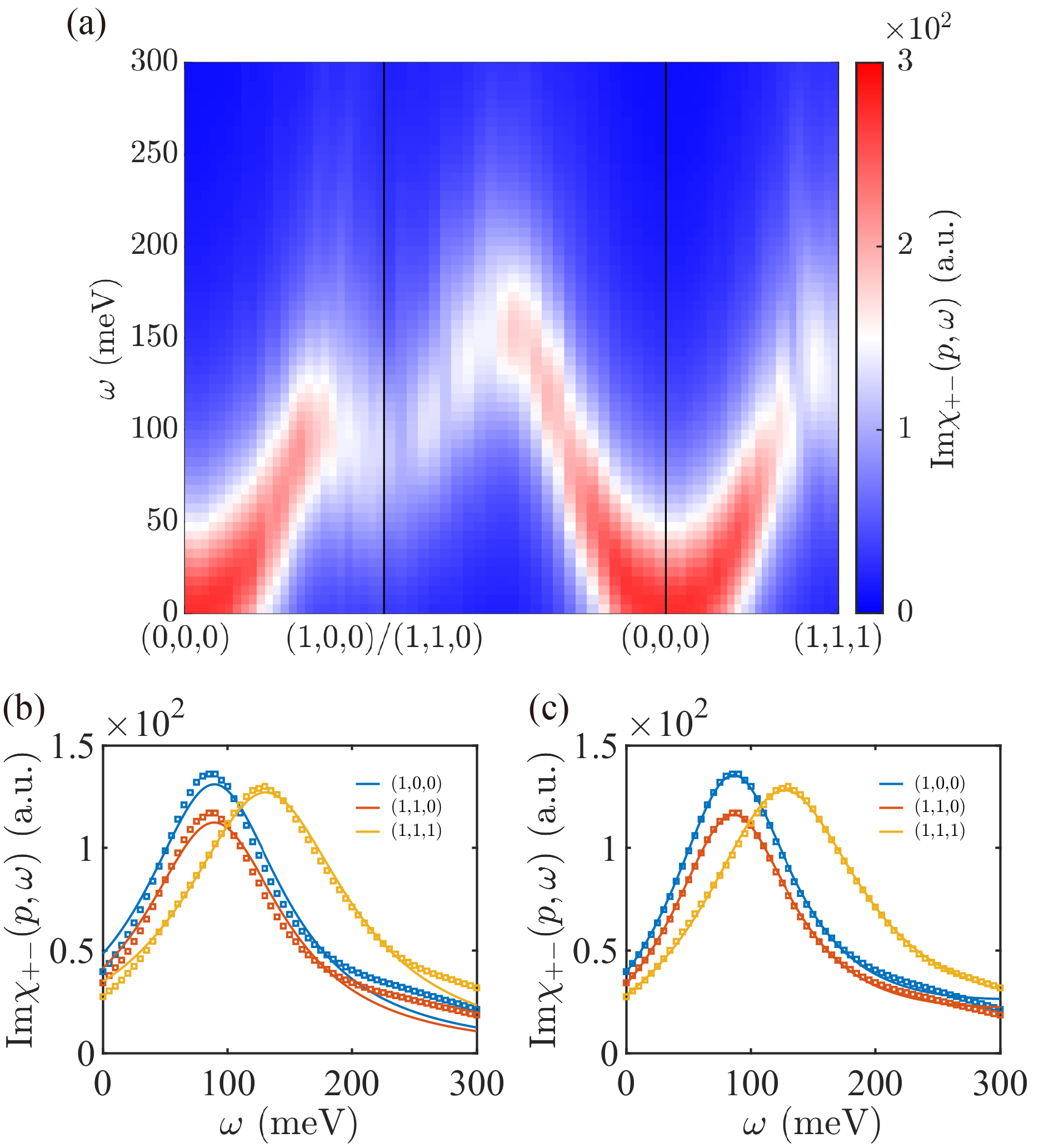}
    \caption{(a) Im$\chi_{\scriptscriptstyle +-}(\bm p,\omega)$ of Cu$_2$MnAl along [100], [110] and [111] directions. Reciprocal lattice vectors of conventional cubic cell are adopted here. (b,c) Im$\chi_{\scriptscriptstyle +-}(\bm p,\omega)$ as a function of $\omega$ for $\bm p = (1,0,0)$, $(1,1,0)$ and $(1,1,1)$. The squares are the calculated data. The solid lines are the fits with (b) symmetric or (c) asymmetric Lorentzian function.}
    \label{fig:Cu2MnAl_Chi}
\end{figure}

\begin{figure}[h]
    \centering
    \includegraphics[width=70 mm]{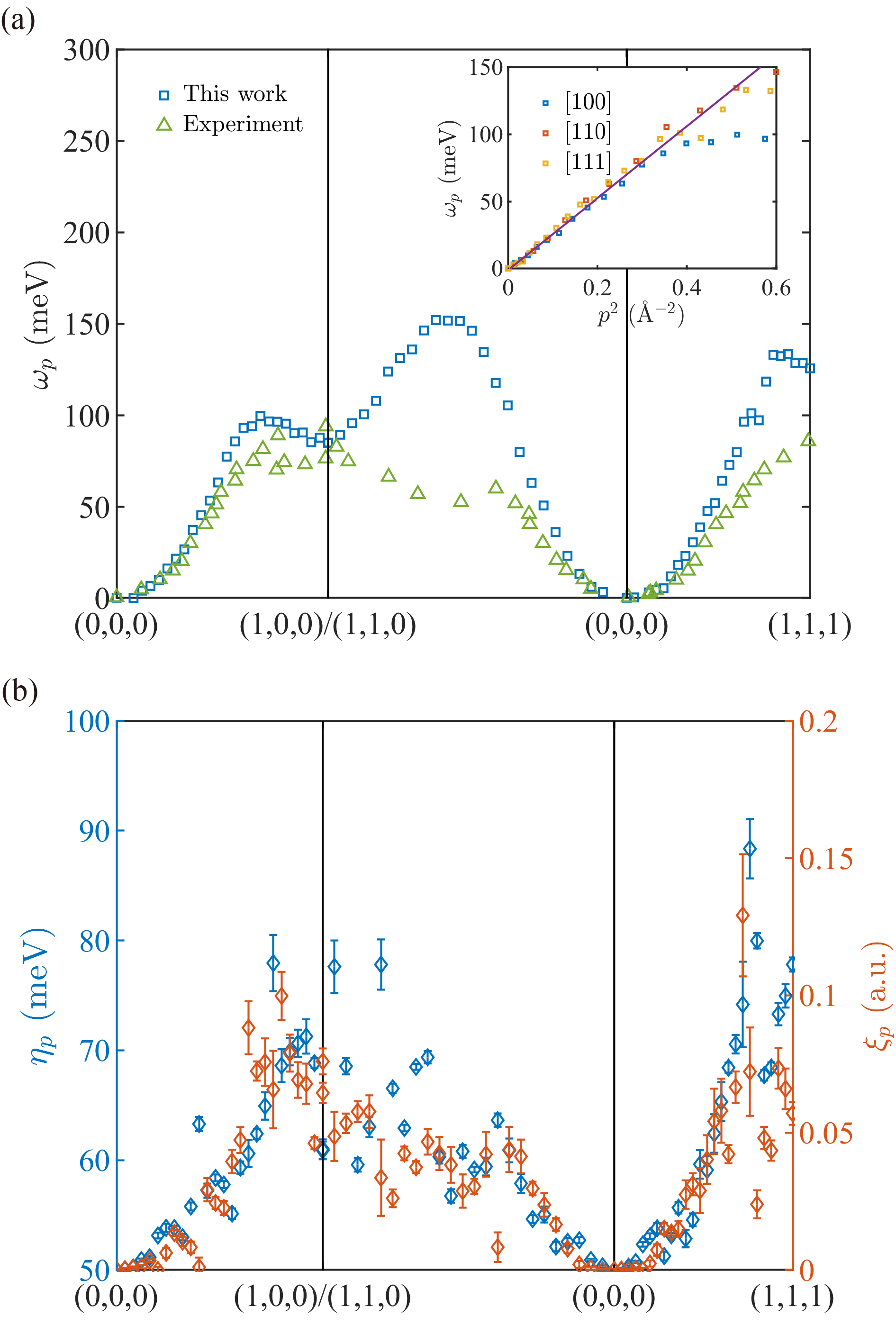}
    \caption{(a) The magnon energy dispersion $\omega_{\bm p}$ of Cu$_2$MnAl along [100], [110] and [111] directions, extracted from the asymmetric Lorentzian fits. $\triangle$ is the experimental data of Tajima et al.~\cite{Tajima1977} Inset is the quadratic fit to the long-wavelength acoustic magnons. (b) Broadening $\eta_{\bm p}$ and asymmetry $\xi_{\bm p}$ of the magnon peaks in Cu$_2$MnAl along [100], [110] and [111] directions. The error bars show 95\% confidence bounds on the fitting parameters from the fits.}
    \label{fig:Cu2MnAl_AsymLorentzFit}
\end{figure}

Fig. \ref{fig:Cu2MnAl_Chi}(a) shows the computed magnon spectral function Im$\chi_{\scriptscriptstyle +-}(\bm p,\omega)$ along the three principal directions. The acoustic magnon branch is seen clearly only at low energies near the Brillouin zone center. The spectral peaks of these low-energy modes can be adequately fitted with the Lorentzian lineshape as in the CrO$_2$ case. The dispersion of the long-wavelength modes is quadratic and isotropic, as demonstrated in the inset of Fig. \ref{fig:Cu2MnAl_AsymLorentzFit} (a). A spin stiffness coefficient $D = 268$ meV$\cdot$\AA$^2$ is procured from the quadratic fit, which is about 1.5 times larger than the experimental value of 175 meV$\cdot$\AA$^2$.~\cite{Tajima1977}

Notably, at higher energies and near Brillouin zone boundaries, the magnon peaks become fuzzier and broader, attesting to substantial Landau damping in this material. In stark contrast to the CrO$_2$ case with almost no Landau damping, the coupling to the Stoner continuum in Cu$_2$MnAl bestows a magnon peak at a given $\bm p$  an asymmetric profile that defies a Lorentzian fit, as shown in Fig. \ref{fig:Cu2MnAl_Chi}(b). The stronger Landau damping in this system is consistent with the absence of the spin-flip gap as shown in Fig.\ref{fig:Cu2MnAl_struc}. Viewing the coupling to the Stoner continuum as a Fano-type resonance,  we superimpose a linear function on the Lorentzian to describe the asymmetric line shape, as
\begin{equation}
A(\bm p,\omega)=\frac{a_{\bm p}\eta_{\bm p}}{(\omega-\omega_{\bm p})^2+\eta_{\bm p}^2}+\xi_{\bm p}(\omega-\omega_{\bm p})
\end{equation}
with $\omega_{\bm p}, a_{\bm p}, \eta_{\bm p}$ and $\xi_{\bm p}$ as fitting parameters. 

As it turns out, this simple modification leads to satisfactory fitting for the entire spectrum, as evidenced in Fig.\ref{fig:Cu2MnAl_Chi}(c). The extracted magnon dispersion is then shown in Fig. \ref{fig:Cu2MnAl_AsymLorentzFit}(a), which coincides with the calculated results of Buczek et al.~\cite{Buczek2009,BuczekThesis2009} and agrees well with the experimental observations along [100] direction.~\cite{Tajima1977} Along the [110] and [111] directions where the Landau damping seems more pronounced, our computed dispersion shows significant discrepancy from the experimental one. For low-energy modes, $\eta_{\bm p}$ is dominated by the artificial broadening parameter $\eta$ and the asymmetry is small, as shown in Fig. \ref{fig:Cu2MnAl_AsymLorentzFit}(b). With increasing energies, however, the broadening quickly exceeds $\eta$ and the asymmetry becomes pronounced, especially near Brillouin zone boundaries, both providing quantitative characterization of the Landau damping.

\section{Summary and outlook}

In conclusion, we report an implementation of the DFPT method in the PAW framework, which is capable of computing the full linear susceptibilities of real materials. A nested iterative procedure is employed to self-consistently solve the Sternheimer equations, to procure linear susceptibilities along with the first-order wavefunctions and densities in monochromatic and periodic external electromagnetic fields. The time cost of each DFPT calculation (given an external field direction, momentum and frequency) is comparable to that of a corresponding Kohn-Sham DFT calculation.

As a demonstration, we compute the spin wave spectra for CrO$_2$ and Cu$_2$MnAl. Gapless magnon dispersion is obtained for both materials from the calculations without spin-orbit coupling. The spin stiffness coefficient extracted from the quadratic fit is in agreement with experimental value for CrO$_2$ but 1.5 times larger for Cu$_2$MnAl. The Landau damping in CrO$_2$ is insignificant due to its half-metallic nature, while in Cu$_2$MnAl is remarkable at high energies and can be quantified with a simple asymmetric Lorentzian fit. LSDA+U method as well as the effect of spin-orbit coupling are examined in CrO$_2$, from which the former highly overestimates the magnon energy, while the latter gives rise to a Goldstone gap quadratic in spin-orbit coupling strength $\lambda$.

There is clearly room for future developments, to make the current implementation more efficient and versatile. From an algorithm viewpoint, the occupied subspace is not projected out in Sternheimer equations in the current implementation. As the contribution to the first-order wavefunctions from the occupied states does not contribute to the first-order densities, projecting out the occupied subspace~\cite{Baroni2001} can potentially improve the efficiency and stability of the nested iteration. As an additional benefit of the projection, it also renders the principle integrals explicit and amenable to analytic techniques, which can further reduce the number of $\bm k$-points required and improve efficiencies. Alternative iterative techniques should be tested in general, for both inner and outer loops, especially in conjunction with the projection.

From a physics viewpoint, a few tasks are on immediate agenda and new possibilities are clearly on the horizon, beyond the initial demonstrations presented herein. For the spin-wave spectral functions, it will be valuable to compare the computed spectra with experimental results for more materials. A particularly interesting comparison can be made between the dispersion relations obtained \textit{ab initio} from our DFPT implementation and those from Heisenberg models parametrized from constrained DFT energies on the basis of the magnetic force theorem.~\cite{Liechtenstein1984,liechtenstein1987,Bruno2003} Such comparisons should be examined in detail for materials in the localized and the itinerant limits, as well as for the continuum falling in between. Further systematic studies for the gradient correction (as in generalized gradient approximations)  and for the Hubbard correction in LSDA+U method can reveal the effect of correlation on the spin-wave spectra. As the first-order wavefunctions are also produced in our code, it is also tempting to evaluate other physical properties, related to density and current responses, such as the magnetoelectric coupling and related transport coefficients. A particular connection may be made by observing that
\begin{equation}
    W = f + f\chi f
\end{equation}
is the screened kernel, which now includes the charge, spin and cross screening effects. This will  enable analyzing the many-electron effects in magnetic materials with strong spin-orbit coupling, and potentially evaluating novel bound states from the screened charge/spin interactions.

\begin{acknowledgments}
    We acknowledge the financial support from the National Natural Science Foundation of China (Grant No. 11725415), the National Key R\&D Program of China (Grant Nos. 2018YFA0305601 and 2021YFA1400100), and the Innovation Program for Quantum Science and Technology (Grant No. 2021ZD0302600).
\end{acknowledgments}

%\renewcommand{\theequation}{A-\arabic{equation}}
%  % redefine the command that creates the equation no.
%\setcounter{equation}{0}  % reset counter 
%\section*{APPENDIX}  % use *-form to suppress numbering

% \subsection*{A.1 Cotangent bundle}

%\bibliographystyle{plain}
%\bibliography{dfpt}


\end{document}